# GUI-based Pedicle Screw Planning on Fluoroscopic Images Utilizing Vertebral Segmentation

Vivek Maik
*Healthcare Technology Innovation Centre*
*Indian Institute of Technology, Madras*
maik.vivek@htic.iitm.ac.in

Aparna Purayath
*Healthcare Technology Innovation Centre*
*Indian Institute of Technology, Madras*
aparna_p@htic.iitm.ac.in

Durga R
*Healthcare Technology Innovation Centre*
*Indian Institute of Technology, Madras*
durga.r@htic.iitm.ac.in

Manojkumar Lakshmanan
*Healthcare Technology Innovation Centre*
*Indian Institute of Technology, Madras*
manoj@htic.iitm.ac.in

Mohanasankar Sivaprakasam
*Department of Electrical Engineering*
*Indian Institute of Technology, Madras*
Chennai, India
mohan@ee.iitm.ac.in

*Abstract*—The proposed work establishes a novel Graphical User Interface (GUI) framework, primarily designed for intraoperative pedicle screw planning. Current planning workflow in Image Guided Surgeries primarily relies on pre-operative CT planning. Intraoperative CT planning can be time-consuming and expensive and thus is not a common practice. In situations where efficiency and cost-effectiveness are paramount, planning to utilize fluoroscopic images acquired for image registration emerges as the optimal choice. The methodology proposed in this study employs a simulated 3D pedicle screw to calculate its coronal and sagittal projections for pedicle screw planning using anterior-posterior (AP) and lateral (LP) images. The initialization and placement of pedicle screw is computed by utilizing the bounding box of vertebral segmentation, which is obtained by the application of enhanced YOLOv5. The GUI front end includes functionality that allows surgeons or medical practitioners to efficiently choose, set up, and dynamically maneuver the pedicle screw on AP and LP images. This is based on a novel feature called synchronous planning, which involves correlating pedicle screws from the coronal and sagittal planes. This correlation utilizes projective correspondence to ensure that any movement of the pedicle screw in either the AP or LP image will be reflected in the other image. The proposed GUI framework is a time-efficient and cost-effective tool for synchronizing and planning the movement of pedicle screws during intraoperative surgical procedures.

## I. INTRODUCTION

Recent methods for pedicle screw placement, incorporating freehand techniques, fluoroscopy-guided, computed tomography (CT) guided procedures, and robotic assistance have been employed over the last few decades. However, no single approach has successfully integrated the heightened precision of image-guided surgery with the reduced preoperative planning time characteristic of the freehand technique [1]. However, even the slightest error in placment of pedicle screws could result in serious impediment, such as neural damage, dural tearing, and potential injury to vascular or visceral structures [2]. Thus there is a necessity for better dependable and reliable pedicle screw insertion techniques with direct dependence on computation techniques and intraoperative imaging. [3]. The plan for screw insertion involves evaluating the morphology of the vertebra and analyzing pedicular dimensions, such as length, radius, angular measurements leading to the computation of entry, centra and target points on the pedical screw. The optimal trajectory would then be along the simulated cylindrical axis of the pedicle screw. [4] [5]. Inadequate planning at both preoperative and intraoperative stages can nullify the advantages offered by an intraoperative navigation system. Therefore, it is crucial to meticulously plan procedures to enhance patient outcomes. Currently, the gold standard imaging modality for preoperative planning in most pedicle screw placement cases, particularly in trauma situations is computed tomography (CT) [6]. Various crucial factors should be considered for preoperative planning when utilizing navigation, including patient positioning, bed selection, placement of digital reference arrays, and the operating room setup. Additionally, consideration should be given to the patient's habitus and the specific dimensions of the image gantry [7]. The primary limitation of CT-based planning is the altered patient position, transitioning from the supine to prone position between preoperative CT scans and intraoperative procedures. [2] In situations where preoperative CT is unnecessary, and image-guided navigation is employed, intraoperative planning emerges as the optimal choice. The planning of pedicle screw trajectories using a navigated probe enables precise placement but requires an O-Arm or intraoperative CT system [8]. The impracticality of intraoperative CT, attributed to its cost, console space requirements, and radiation exposure, prompts the consideration of alternative imaging modalities like fluoroscopy for planning insertions. This method utilizes the images acquired solely for patient registration without the need for additional imaging.

The GUI front end features interactive capability that allows surgeons or medical practitioners to quickly select, set up,

and dynamically move the pedicle screw on AP and LP images. This is based on synchronous planning, a novel feature that includes correlating pedicle screws in the coronal and sagittal planes. One of the primary benefits of synchronous planning with our GUI is that surgeons only need to plan pedicle screws on either AP or LP images. Synchronous correspondence assures that AP planning in the AP image is computationally translated to the LP image as LP planning, and vice versa. This eliminates the tedious job of manually matching AP and LP planning for high precision, which would be required if the suggested GUI interface did not facilitate synchronous planning. Existing image-based GUI planning for pedicle screws relies heavily on 3D modalities like CT or MRI. The suggested GUI-based pedicle screw planning is not dependent on 3D data because it just requires 2D flouroscopic AP and LP images. The 2D fluoroscopy image is also the only imaging modality that can be acquired intra-operatively, whereas the 3D modality can only be acquired pre-operatively, necessitating intraoperative registration prior to pedicle screw planning. We do not need a registration process with the proposed 2D pedicle screw planning, and surgeons are given intra-operative real-time data to pedicle screw planning, which includes automatic screw initialization and projective correspondence-based plan synchronization via a GUI, which is a significant advantage of over other methods that emphasize 3D modalities.

This paper is organized as follows: Section II provides an overview of the current computer-assisted planning methods. It also summarises deep learning-based vertebral segmentation methods. Section III describes the methodology devised for the proposed work. Section IV presents the observations and discussions on the framework. Section V encompasses the conclusions and future work.

## II. RELATED WORK

Many insertion planning procedures is typically based on manual select and drop method on an intraoperative image, showcased on a visualization workstation. This often requires repeated iterations to achieve satisfactory outcome, can potentially disrupt the workflow and introduce human errors. As a result, researchers have come up with computer-assisted planning methods to overcome the drawbacks in manual screw selection procedures. [9]. Wicker and Tedla [10] and Lee et al [11] devised the optimal screw trajectory through the extraction of the pedicle region from user input and subsequent reconstruction of the pedicle shape. On the other hand, Soliman et al [12], and Caprara et al [13] derived bone mechanics and mineral density characteristics from flouroscoping images and determined the pedical screw trajectory. [9].

The advent of Deep Learning (DL) has brought about a transformative shift [14] [15], particularly in the Healthcare industry. The implementation of DL techniques has successfully automated formerly manual and labor-intensive processes, presenting significant advancements. Xiaozhi Qi et al [16] developed an automated algorithm for pedicle screw planning through segmentation of the patient's spine using preoperative CT scans. Initially, each vertebra is segmented using a trained network model, from which a local coordinate system is derived proportional to its anatomical characteristics. Subsequently, the identification of feature points in the images allows for the automatic identification of the screw placement region as well as the trajectory path for screw placement, streamlining the surgical path planning process. Lisa et al [17] devised an automated screw positioning within the CT volume based on vertebra instances using patches. This approach utilized a U-Net framework for extracting the region indicating the location of the screw pair. The majority of DL techniques have been employed on CT volumes to automate planning. Though some DL techniques have been applied to fluoroscopic images, their usage has primarily focused on post-operative analysis of the procedure. Hooman Esfandiari et al [18] proposed a convolutional neural network (CNN) for pedicle screw segmentation in X-Ray projections for verification of accurate screw insertion.

Vertebral segmentation methods on fluoroscopic images have been explored for varying applications such as Cobb angle measurement [19] measurement of vertebral angular movements, [20] or deformity identification [21]. A vertebrae segmentation method, proposed by Lecron et.al [20] employs parallel computing for analyzing heterogeneous topologies. This approach achieves improved accuracy, and diminishes the cost associated with data transfer between memory units, thus rendering it suitable for managing extensive datasets. Mushtaq et al [21] developed a system for the localization and segmentation of lumbar vertebrae from sagittal fluoroscopic images for identifying spinal deformities. Based on the localization and vertebra of interest, a mask is generated. A HED U-Net model is employed to identify edges with the image and the mask as inputs. Nevertheless, a limitation of this approach is its lack of consideration for the coronal plane and the synchronization between the two planes.

For the vertebral segmentaion, our proposed approach relies on the enhanced YOLOv5 module with Adaptive Fusion and Co-ordinate Attention for precise vertebrae localization in 2D fluoroscopic images.

## III. METHODOLOGY

### A. GUI Design and Working

The GUI framework is well-suited for semi-automatic screw planning and placing in the spine using vertebrae segmentation. The GUI is a comprehensive tool that helps surgeons with intraoperative planning and guidance for pedicle screw insertion. The workflow begins with the acquisition of fluoroscopic images of the spine from both AP and LP views. Subsequently, these images undergo pre-processing and augmentation to enhance quality and diversity. The annotated images form the training dataset for an enhanced YOLOv5, which is then trained to segment vertebrae in the fluoroscopic images. The trained model is employed for the inference and prediction of new fluoroscopic images which can then be used in the surgical phase; the surgeon selects patient-specific AP and LP images. The GUI displays coronal and sagittal view projections of the spine, allowing the surgeon to select the vertebrae of interest from which the enhanced YOLOv5 defines a bounding box for screw placement. Utilizing projective correspondence, screws are added to the bounding box such that the planning of the AP screw will reflect on the LP and vice versa as shown in Figure(1). The GUI facilitates a simulation of the surgical screws in the patient's spine, allowing the surgeon to adjust placement as necessary; the GUI generates a surgical plan detailing screw size, type, and location.

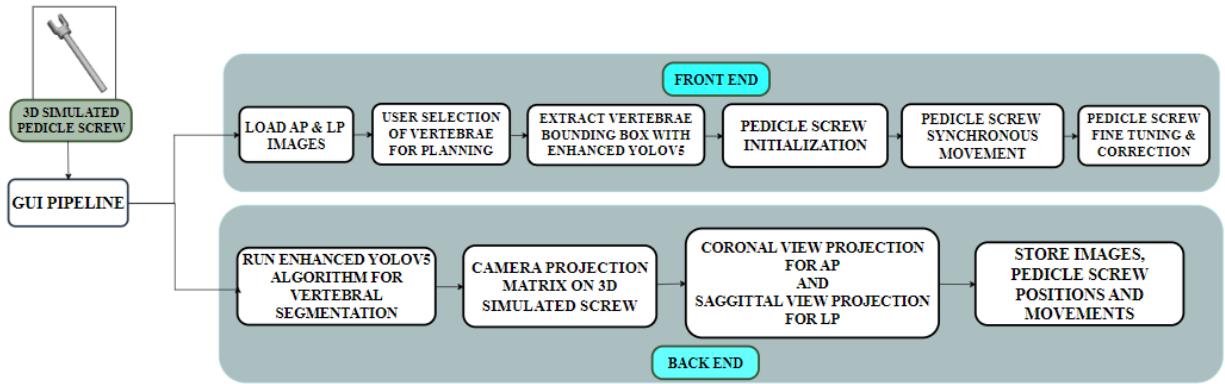

Fig. 1. Block diagram of proposed work: graphical user interface (GUI) for planning and installing screws in vertebrae using 3D segmentation

Following the pre-processing stage, which involves Contrast Limited Adaptive Histogram Equalization (CLAHE) and Non-Local Means Denoising, we deploy an enhanced YOLOv5 model to accurately locate vertebrae in AP and LP images. The vertebral detection technique based on bounding boxes utilizes training labels and incorporates Adaptive Fusion and Co-ordinate Attention modules (AF-CA) to enhance the performance of YOLOv5 in handling intricate features, such as spine vertebrae. The vertebral detection technique based on bounding boxes utilizes training labels and incorporates Adaptive Fusion and Co-ordinate Attention modules (AF-CA) to enhance the performance of YOLOv5 in handling intricate features, such as spine vertebrae. The incorporation of the AF module aggregates local features from two different scales, followed by interpolation and feature fusion, like Spatial Pyramid Pooling (SPP), facilitating the detection of objects at various scales and capturing spatial information. This is crucial for fluoroscopic images with varying contours in shape and size. Test results indicate that the AF module significantly aids in identifying smaller, intricate vertebrae in spine fluoroscopic images. The CA module can capture input features along the vertical and horizontal directions. These feature maps are then multiplied with the input feature map, and this provides for better object localization. This study primarily focuses on examining the synchronized planning of pedicle screws between AP and LP images. Consequently, it does not provide a detailed explanation of the augmented YOLOv5 method employed for obtaining the vertebral anatomical bounding box. The graphical user interface (GUI) functions by projecting a 3D simulated pedicle screw design into the fluoroscopic image space using existing calibration data. Once the 3D screw is introduced into the fluoroscopic image space, we capture the coronal and sagittal projections of the screws. The resulting 2D projection screw graphs from the 3D simulated pedicle screw are superimposed on AP and LP images, respectively.

Establishing correspondence between the projection plots enables the computation and application of any movement in the AP 2D (AP screw) screw to the LP 2D screw (LP screw), and vice versa. This synchronization of movement in both AP and LP 2D screws is achievable when the movement is applied to a 3D pedicle screw. A noteworthy innovation in this work is the synchronization of planning in AP with planning in LP, along with the synchronization of screw movement in both 2D and 3D. This GUI-based screw synchronization proves highly advantageous in surgical planning, allowing for the automatic computation and application of planning in various aspects, such as the location, translation, or rotation of pedicle screws in either 3D or 2D. The GUI facilitates the selection of vertebrae by surgeons through mouse pointers, initiating the screw placement process automatically using the improved YOLOv5 bounding box. Subsequently, the GUI enables the manipulation of the initialized screw through mouse control, facilitating the synchronization of screw planning between AP and LP.

### B. GUI Based Pedicle Screw Planning

The GUI is employed to generate a 3D pedicle screw/cylinder object shape, which is then presented in 2D

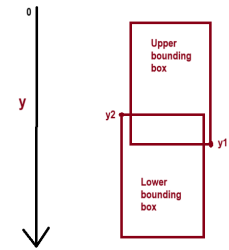

Fig. 2. Overlapping vertebrae bounding boxes for computation of pedicle screw entry and target point

space through two plots superimposed on AP and LP images in the (x, y) domain. The overlay on the LP plot assumes a sagittal projection, considering the LP image is positioned beside the 3D simulated pedicle screw. Conversely, the AP plot overlay estimates the coronal projection, assuming the AP image is located beneath the 3D simulated pedicle screw. The overlapping region of interest is identified from these coronal and sagittal projections, serving as the basis for synchronizing pedicle screw planning between AP and LP images. In simpler terms, surgeons can plan on AP, and the planning will be reflected in the LP image. Establishing an accurately planned correspondence between AP and LP is expected to be more efficient than a GUI design emphasizing separate planning on AP and LP. The target point denotes the final location for screw implantation in the 3D simulated screw, while the entry point signifies the incision site for screw placement. In this study, we determine the bounding box coordinates of the vertebrae using improved YOLOv5 segmentation. The surgeon maps the surgical vertebrae of interest to the bounding box of the specific vertebrae once identified. The computation of entry and target points utilizes these bounding box coordinates as input, including the labeling of the vertebral body and upper and lower intervertebral disks. For precise screw placement, the simulated pedicle screw is inserted into the vertebral body region of the spine. To exclude the intervertebral disk length (IVDL) while identifying the entrance and target positions, two methods are employed. The first method involves manually measuring the vertical disk length across various vertebral pictures and subtracting it from the bounding box, effectively removing the IVDL. The second method, based on vertebrae labeling, requires a minimum of two bounding boxes, indicating the detection of at least two vertebrae by the enhanced YOLOv5. These overlapping bounding boxes share a common area, the IVDL, which can be excluded, as illustrated in Figure(2). In the experimental phase, the initial bounding box estimation using enhanced YOLOv5 was trained and tested on a dataset comprising 578 images obtained from cadavers and lumbar phantoms. This dataset was partitioned into 500 training images, 58 validation images, and 20 test images, with a specific focus on improving the YOLOv5 algorithm for vertebrae segmentation.

The coronal plane vertebra's bounding box is split in half vertically. The pedicle screw is positioned in the left half of the bounding box and the computation is carried out in the left portion of the box when the input position is "L" (left). In the same way, the computation for "R" (right) is carried out in the bounding box's right section. The screw is shown horizontally along the Sagittal plane vertebra's bounding box, regardless of whether it is on the left or right. Points are therefore determined using the bounding box's sides Consider the coordinates of the target point, C1, as (X1, Y1, Z1), and the entry point, C2, as (X2, Y2, Z2). Let AP_R represent the selected vertebra bounding box in the coronal plane plot, and LP_R denote the selected vertebra bounding box in the sagittal plane plot. The Y and Z coordinates of C1 and C2 are computed from AP_R, while the X coordinate is calculated from LP_R. Additionally, a fixed padding of 'pad1' is applied for C1, and 'pad2' is used for C2, as depicted in Figure(3).

The segmented vertebrae output serves as the basis for computing the entry and target coordinates (C1 and C2), as explained earlier. This process enables the initialization of the 3D simulated pedicle screw in the shape of a cylinder in both AP and LP images, with the screws capable of simultaneous movement and adjustment. Following the initialization of the cylinders, surgeons can manipulate them by choosing a circular or longitudinal area using a mouse. Figure(4) illustrates that the longitudinal area selection corresponds to the translational movement of the cylinder, while the circular area selection allows for resizing and rotation of the cylinder. The movement of the cylinder along circular and longitudinal planes captures and records the motion, ensuring that this action in the AP plane is mirrored on the lateral image, and conversely. This synchronization occurs as the cylinder is generated based on entry and target locations. With each movement, the cylinder computes and updates the tangent line, vector, and circle points corresponding to the screws in both the AP and LP images.

## IV. RESULTS AND DISCUSSIONS

The GUI interface provides options to load both AP and LP images, align them with the same plane rotation, and zoom into the images. The right-side GUI panel offers the capability to select the desired vertebrae, such as the Lumbar (L3-L5) for surgeons. Given that a single fluoroscopic image can capture at most 2 to 3 full vertebrae, the GUI accommodates a maximum of 3 vertebrae for initialization based on user-selected points. Surgeon labels L4 & L5 (visualized as blue points on the image) and the labeled vertebrae are highlighted in green on the GUI panel. Each vertebra requires the surgeon to choose two points, one for AP and another for LP, with each point positioned at the center of the vertebrae. Figure(5) provides a visual representation of the GUI interface for pedicle screw planning. The GUI was tested for image-based screw placement and planning. From the user-selected vertebrae points, the enhanced YOLOv5-driven background model produces a bounding box, as depicted in Figure(6). Utilizing the bounding box output, the initialization of pedicle screw parameters for projecting the 3D screw onto AP and LP images is determined,

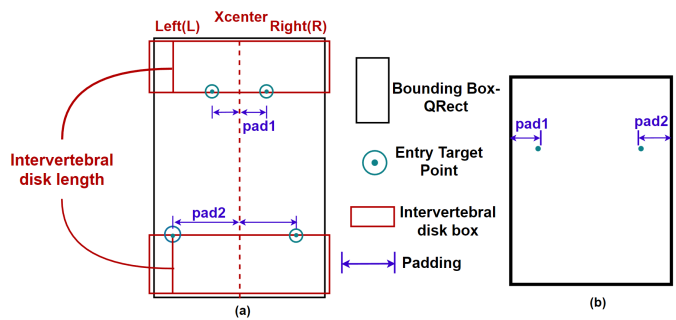

Fig. 3. Computation of entry and target point positions on the Coronal plane & Sagittal plane plots (a) AP_R, (b) LP_R

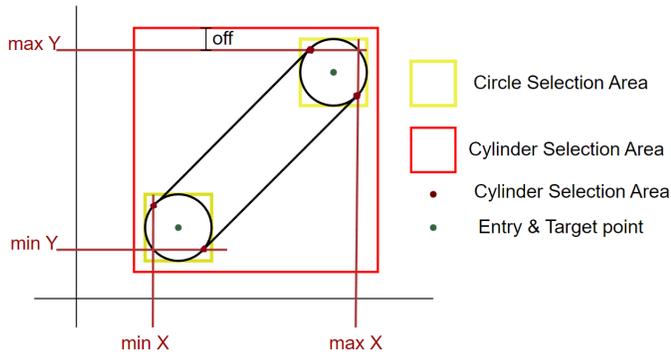

Fig. 4. Cylinder selection area

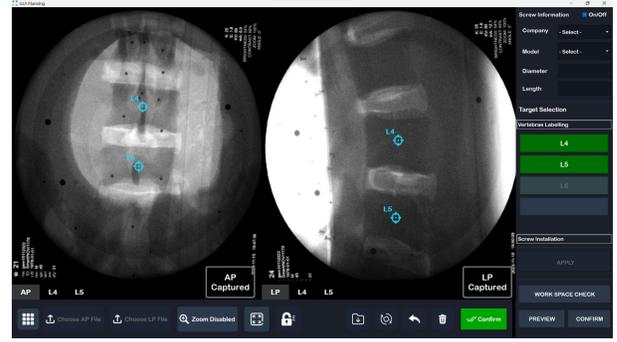

Fig. 5. Simulated Pedicle Screw Planning GUI interface

**Algorithm 1** Computation of Entry and Target Points

**Require:** APR: AP bounding box
  LPR: LP bounding box
  IVDL: Intervertebrae Disk Length
**Ensure:** C1(X1,Y1, Z1): Entry Point
  C2(X2,Y2, Z2): Target Point

**Padding Inputs:**

  C1pad ← val + 1 + Left
  C2pad ← val1 - Right

**Box Center Point:**

  Xcenter ← APR.Xtopleft + APR.Width/2

**C1 and C2 Computation:**

  Y1 ← Xcenter + val pad1
  Y2 ← Xcenter + (val - (APR.width()/2 - pad2))
  Z1 ← APR.YtopLeft + IVDL
  Z2 ← APR.Ybottomright - IVDL
  X1 ← LP.left + pad1
  X2 ← LP.right - pad2 =0

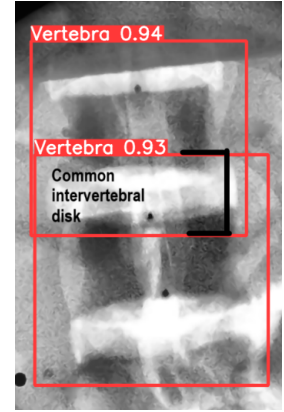

Fig. 6. Enhanced YOLOv5 will determine the vertebral bounding box as shown here

as illustrated in Figure(7). The screw planning GUI option empowers the surgeon to document the screw movements in AP and LP images. Specifically, (L4-L6)-L records movements in the AP images, while (L4-L6)-R records movements in the LP images. The GUI is also simultaneously based on the simulated 2D screw-on image to compute the diameter and length of the screw which could be handy for the surgeons in selecting the real screws.z

Achieving synchronization of the screw's position in both coronal and sagittal plane plots is essential for accurate 3D placement. Discrepancies caused by imaging variations are addressed by computing mismatch along the common dimension between target and entrance locations in both the AP and LP image planes and comparing them to ground truth differences. These discrepancies are then removed from the common dimension values to ensure that the screw aligns exactly with the surgeon's intended position. Continuous movement is facilitated by updating entry and target points based on the surgeon's inputs, allowing the cylinder to vary in direction in response to the user's motions throughout the plot.

## V. CONCLUSION

The proposed GUI-based setup for intraoperative pedicle screw planning makes significant contributions: (i) It provides an easily usable interface framework for surgeons and practitioners, streamlining the planning process. (ii) The GUI automatically establishes a computational correspondence between anterior-posterior (AP) and lateral (LP) images. Planning only needs to be done on one image, as it will be automatically updated on the other image. (iii) Through the use of the back-end enhanced YOLOv5, the GUI interface obtains vertebral segmentation, which is then utilized to initialize the positions of pedicle screws in the respective images. (iv) Discrepancy errors in the initialization process, arising from C-Arm orientation and distortion, are effectively compensated for during the initialization process.

Using the proposed GUI framework, pedicle screw planning can be performed intraoperatively using conventional fluoroscopic imaging in a time-efficient and cost-effective manner. Unlike the following stated by Ulf Bertman et al [22], that planning through fluoroscopy does not offer extended functionality, this GUI allows surgeons to plan pedicle screw insertions by accurately determining the diameter and length of the screws. Further work involves testing the GUI framework in various clinical scenarios utilizing the lumbar phantom. Clinical validation is essential, and it will be conducted by assessing the plan versus position error in pedicle screw planning using the GUI.

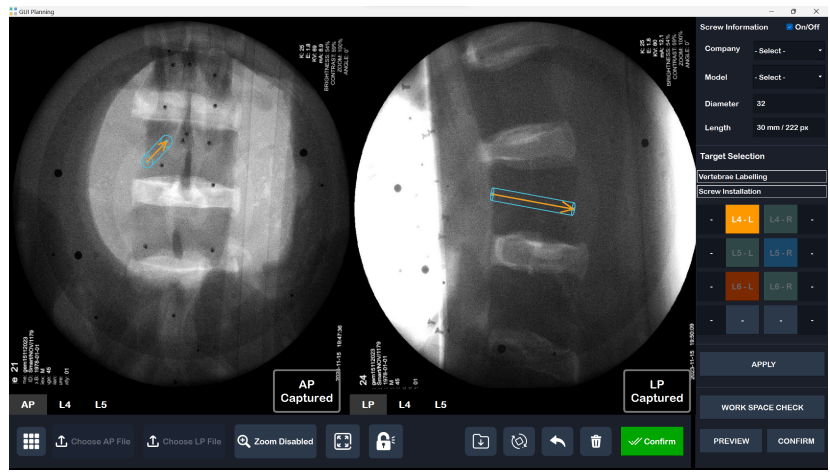

Fig. 7. Synchronous Pedicle Screw Placement and Planning on AP and LP images